# p/e GEOMETRIC MASS RATIO

**Gustavo González-Martín**

Departamento de Física, Universidad Simón Bolívar,

Apartado 89000, Caracas 1080-A, Venezuela.

Web page URL  http:\\prof.usb.ve\ggonzalm\

A previously proposed geometric definition of mass in terms of energy, in a geometrical unified theory, is used to obtain a numerical expression for a ratio of masses of geometrical excitations. The resultant geometric ratio is approximately equal the ratio of the proton to electron physical masses.

# 1. Introduction.

We have presented a definition of mass [1], within our geometric unified theory of gravitation and other interactions, in terms of the concept of self energy of the non linear self interaction in a geometric space, leading to the mass term in Dirac's equation. The concept of mass plays a fundamental role in relativity as shown by the relation between mass and energy and the principle of equivalence between inertial and gravitational mass. It is then of interest to find a method to carry a numerical calculation to test the proposed geometric model.

On the other hand we also have considered an approximation to the geometric non linear theory [2] where the microscopic physical objects (geometric particles) are realized as linear geometric excitations, geometrically described in a jet bundle formalism shown to lead to the standard quantum field theory techniques. These geometric excitations are essentially perturbations around a non linear geometric background space, where the excitations may be considered to evolve with time. In this framework, a geometric particle is acted upon by the background and is never really free except in absolute empty background space (zero background curvature). The background space carries the universal inertial properties which should be consistent with the ideas of Mach [3] and Einstein [4] that assign fundamental importance of far-away matter in determining the inertial properties of local matter. We may interpret the geometric excitations as geometric particles and the background as the particle vacuum. This is a generalization of what is normally done in quantum field theory when particles are interpreted as vacuum excitations. The vacuum is replaced by a geometric symmetric curved background space.

In quantum electrodynamis the self energy is a source of problems. The infinite energy of the vacuum state is resolved by subtracting the infinite self energy by using the ordered products of operators. Similarly, the electron mass is infinite in quantum electrodynamics. This problem is resolved by subtracting the infinite self energy effect and assuming the experimental value for the mass, in the process of renormalization.

These problems do not arise in our treatment. If we consider geometric excitations on a background, the expression defining mass may be expanded in a perturbation series around the background with a *finite zeroth order term given entirely by the system self energy in terms of the background current and connection*. The higher order terms are corrections depending on the excitation self interaction. As indicated in previous work, these corrections correspond to a geometric quantum field theory and may be taken care by the standard quantum field techinques. There is no ad-hoc assumption on our calculation because it is based, as indicated in the first paragraph, in a definition of mass given in a unified theory long before this calculation was attempted.

In this article we limit ourselves to the zeroth order background contribution, which we consider a bare mass, subject to further corrections. The existence of a simple particular solution to the background field equations allows the possibility of calculating these geometrical bare masses and obtain mass ratios using ratios of volumes of symmetric spaces, obtained from the structure group as cosets with respect to some of its subgroups. It should be noted that the structure group $G$ of the theory, SL(4,R), has been used to describe particle properties [5,6,7] in another approach. First we shall review the concept of geometric excitations.

The fundamental dynamic process in the theory is the action of the connection on the frame. Since the connection is valued in the Lie algebra of $G$ and the frame is an element of the group $G$, the dynamics is realized by the action of the group on itself. The principal bundle structure of the group, $(G,K,L)$, provides a natural geometric interpretation of its action on itself. A particular subgroup $L$ defines a symmetric space $K$, the coset $G/L$, the base space of the bundle. The subgroup $L$, the fiber of the bundle, is the isotropy subgroup of the coset $K$ acts on itself. The complementary coset elements act as translations on the symmetric coset.

This geometric interpretation may be transformed into a physical interpretation if we choose $L$ to be homomorphic to the spinor group, SL(2,C), related to the Lorentz group. The action of $L$ is then interpreted as a pseudo rotation (Lorentz transformation) of the *external* space, the tangent space $TM$ of the physical space time manifold $M$, defining a metric. The action of the complementary coset $K$ is interpreted as a translation in an *internal* space, the symmetric coset $K$ itself. There is then a non trivial geometric relation between the internal and external spaces determined by the Clifford algebra structure of the manifold. The space $K$ is the exponentiation of the odd sector of the Clifford algebra and is related to local copies of the tangent space $TM$ and its dual cotangent space $T^*M$ and may be interpreted as a generalized momentum space. The states of momentum $k$ would correspond to a point of $K$.

It follows that the frame excitations are also acted by the connection and evolve as representations of $G$. Different observers would measure different relative momenta $k$ for a given excitation. A measurement for each $k$ corresponds to a function in momentum space. An abstract excitation is an equivalence class of these functions, under the relativity group. Since the group space itself carries its own representations, the realization of excitations as representations defined on the group space have a fundamental geometric character. The geometric action of the $K$ sector is a trans-lation on itself and the functions on *K* are the observable internal linear representations. The action of the *L* sector is a Lorentz transformation and sl(2,C) spinors are the observable external representations. For these reasons, we *must represent physically observable excitations by classes of spinor valued functions on the symmetric K space*. In particular, we realize them on a vector bundle, associated to the principal bundle (*E,M,G*), with fiber the sl(2,C) representation valued functions on the symmetric *K* space. In mathematics, these representations are called representations of SL(4,R) induced from SL(2,C). Essentially, this is, in fact, done in particle physics when considering representations of the Poincare group.

From our geometric point of view, it has been indicated [8] that the three stable particles, proton, electron and neutrino appear to be representations of the only three dynamical subgroups of SL(4,R) induced respectively from the even subgroups $SL_1(2,C)$ and SL(2,C). This is a generalization of particles as representations of the Poincare group induced from its Lorentz subgroup. Therefore we shall apply the proposed definition of mass to these representations, thus obtaining a physical and mathematically different calculation. It should be noted that there is no contradiction in this calculation with present physical theories, which may be considered as effective theories derived under certain conditions and limits from other theories.

## 2. Bare Masses.

The definition of the mass parameter *m*, in terms of a connection $\omega$ on the principal fiber bundle *(E, M, G)*, has been given in the fundamental defining representation of SL(4,R) in terms of *4×4* matrices, but in general, may be written for other representations using the Cartan-Killing metric $g_C$, defined by the trace. The definition of this metric may be extended to the Clifford algebra *A*, which is an enveloping algebra of both sl(4,R) and sp(4,R). The Clifford algebra *A* is a representation and a subalgebra of the universal enveloping algebra *U* of these Lie algebras. The dimension of the vector space carrying the group representations is the trace of the representation of the identity in *A*. We may write the definition of the mass parameter in any representation of the algebras sl(4,R) and sp(4,R) and the corresponding representation of the common enveloping Clifford algebra $\mathcal{D}(A)$ as

$$m = \frac{1}{4}\operatorname{tr} J^\mu \Gamma_\mu = \frac{\operatorname{tr} J^\mu \Gamma_\mu}{\operatorname{tr} I_A} \equiv \frac{g_C\left(J^\mu \Gamma_\mu\right)}{g_C\left(I_{\mathcal{D}(A)}\right)} \quad . \tag{2.1}$$

It is known that the Cartan metric depends on the representations, but we shall only apply this expression to find ratios within a particular fixed induced representation of the enveloping algebra.

If we consider geometric excitations on a background, this mass may be expanded as a perturbation around the background in terms of the only small parameter $\alpha$, characterizing the excitation,

$$m = \frac{1}{4}\operatorname{tr}\left(J_0^\mu \Gamma_{0\mu} + \alpha J_1^\mu \Gamma_{0\mu} + \alpha J_0^\mu \Gamma_{1\mu} + O(\alpha^2)\right) \quad , \tag{2.2}$$

indicating that the zeroth order term is given entirely by the background current and connection, with corrections depending on the excitation self interaction. As indicated in previous work [9], these corrections correspond to a geometric quantum field theory. In this article we limit ourselves to the zeroth order term which we consider the bare mass of QFT.

The structure group *G* is SL(4,R) and the even subgroup $G_+$ is $SL_1(2,C)$. The subgroup *L* is the subgroup of $G_+$ with real determinant in other words, Sl(2,C). There is another subgroup *H* in the group chain $G \supset H \supset L$ which is Sp(4,R). The corresponding symmetric spaces and their isomorphisms are discussed in appendix B. We are dealing with two quotients which we shall designate as *C* and *K*,

$$K \equiv \frac{G}{G_+} \cong \frac{SL(4,R)}{SL(2,C) \otimes SO(2)} \cong \frac{SO(3,3)}{SO(3,1) \otimes SO(2)} \quad , \tag{2.3}$$

$$C \equiv \frac{H}{L} \cong \frac{Sp(4,R)}{Sp(2,C)} \cong \frac{SO(3,2)}{SO(3,1)} \quad . \tag{2.4}$$

These groups have a principal bundle structure over the cosets and themselves carry representations. The geometric action of the *K* generators are translations on the coset *K*. The functions on *K* are the natural internal representations.

We shall consider, then, the representations of SL(4,R) and Sp(4,R) induced from the subgroups $SL_1(2,C)$ and SL(2,C) over the symmetric spaces $SL(4,R)/SL_1(2,C)$ and Sp(4,R)/SL(2,C), respectively. The geometric induced representations may be realized as sections of a homogeneous vector bundle $(D, K, \mathcal{D}[SL_1(2,C)], G_+)$ with $SL_1(2,C)$ representations $\mathcal{D}$ as fiber $F$ over the coset $K$ [10]. To the induced representation of SL(4,R) on $D$, there corresponds an induced representation of the enveloping Clifford algebra $A$ on $D$ [11]. Furthermore, to the latter also corresponds a representation of the subgroup Sp(4,R) on $D$. In other words, the vector bundle $D$ carries corresponding representations of $A$, SL(4,R) and Sp(4,R). These three representations are functions on $K$ valued on representations of $SL_1(2,C)$. At each point of the base space $M$ we consider the function space $\Sigma$ of all sections of the homogeneous vector bundle $D$. Define a vector bundle $S \equiv (S, M, \Sigma, G)$, associated to the principal bundle $E$, with fiber the function space $\Sigma$ of sections of $D$. The fiber of $S$ is formed by induced representations of $G$.

There is an induced connection acting, as the adjoint representation of $G$, on the bundle $S$. The connection $\omega$ is represented by matrix operators (Lorentz rotations on $L$ and translations on $K$). The induced connection $\omega$ may be decomposed in terms of a set of basis functions characterized by a parameter $k$, the generalized spherical functions $Y_k$ on the symmetric space [12]. If $K$ were compact, the basis of this function space would be discrete, of infinite dimensions $d$. The components, relative to this basis would be labeled by an infinite number of discrete indices $k$. Expressing the Cartan-Killing metric in the induced representation, we formally would have for the mass parameter, in terms of the $\Gamma$, $J$ components

$$m = \frac{1}{4d} \operatorname{tr} \sum_{k,k'} J_{k'}^{k\mu} \Gamma_{k\mu}^{k'} \quad . \tag{2.5}$$

Since the spaces under discussion are non compact, the discrete indices $k$ that label the components, become continuous labels and the summation in matrix multiplication becomes integration over the continuous parameter $k$. In addition, we are working with 4 dimensional matrices and continuous functions on the cosets, and the Cartan-Killing metric in $A$, expressed by trace and integration, introduces a $4V(K_R)$ dimensionality factor for the common carrier space $D$, giving,

$$m = \frac{1}{4V(A_R)} \operatorname{tr} \int dk \int J(k, k_2) \Gamma(k_2, k) dk_2 \quad . \tag{2.6}$$

where $V(A_R)$ is a volume characterizing the dimension of the continuous representation of $A$. We interpret the value of a function at $k$ as the component with respect to the basis functions $Y(kx)$ of the symmetric space $K$, parametrized by $k$, as usually done in flat space in terms of a Fourier expansion. We may say that there are as many "translations" as points in $K$. It should be noted that these "translations" do not form the well known Abelian translation group.

The $G$-connection on $E$ induces a SO(3,1)-connection on $TM$. The combined action of the connections, under the even subgroup $G_+$, leaves the orthonormal set $\kappa_\mu$ invariant [13], defining a geometric relativistic equivalence relation $R$ in the odd subspace $K$. Each element of the coset is a group element $k$ that corresponds to an odd moving frame. A class of equivalent moving frames $k$ is represented, up to an $SL_1(2,C)$ transformation, by a single rest frame, a point $k_0$. Decomposition of $\Gamma$ and $J$ is into equivalence classes of state functions $Y(kx)$. The dimension of the induced representation space, or number of classes of state functions $Y(kx)$ (independent bases), is the volume of a subspace $K_R \subset K$ of classes (non equivalent) of points. The current and connection components are functions over the coset $K$. Physically the integral represents summation of the connection $\times$ current product, over all inequivalent odd observers, represented by a rest observer.

There is a constant solution [14] for the non linear differential equations (see the appendix) that provides a trivial connection $\varpi$ to the principal fiber bundle $(E,M,G)$. This SL(4,R) valued 1-form on $E$ represents the class of equivalent local connection forms represented by the constant solution. At some particular frame $s$, that we may take as origin of the coset, the local expression for $\varpi$ is the constant

$$s^* \varpi = -m_g \kappa_\alpha dx^\alpha = -m_g J \quad . \tag{2.7}$$

All points of $K$ or $C$ may be reached by the action of a translation by $k$, restricted to the corresponding subgroup, from the origin of the coset. As the reference frame changes from $s$ at the origin to $sk$ at point $k$ of the coset, the local connection form changes

$$k^* s^* \varpi = k^{-1} s^* \varpi k + k^{-1} dk = -m_g J + k^{-1} dk \equiv \Lambda + k^{-1} dk \quad , \tag{2.8}$$

which corresponds to the equivalence class of constant solutions $\varpi$. All $k*s*\varpi$ correspond the same constant solution class $\varpi$, seen in the different frames of the coset.

In the principal bundle the constant connection $\varpi$, combines with the constant $J$ to produce a product constant over the coset $K$, as long as the transformation is orthogonal to $J$ [15]. As indicated before, the mass variation, produced by the last term in the equation due to an arbitrary choice of frame, corresponds to inertial effects. The non inertial effects are due to the first term in the right hand side of the equation which corresponds to the symbol $\Lambda$. It is clear that its subtraction from the connection transforms as a connection, because the current $J$ is a tensorial form, and corresponds to the inertial connection. The latter connection is expressed by the last term in the equation and only has a $k$ dependence. The physical contribution to the bare mass parameter may be calculated in the special frame $s$ giving,

$$g_C(s^*\omega \cdot J) = g_C(J^\mu \Lambda_\mu) = -m_g g_C(J^\mu J_\mu) \; , \qquad (2.9)$$

and defining an invariant expression in terms of $\Lambda$, valid for a given representation.

We are interested in the induced representations of SL(4,R) and Sp(4,R) corresponding to the *same* trivial constant solution. In the defining representation of 4 dimensional matrices, the product $JJ$,

$$J_G \cdot J_G = e^{-1}\kappa_\mu e e^{-1}\kappa^\mu e = -4I = J_H \cdot J_H = J'_G \cdot J'_G = J'_H \cdot J'_H \; , \qquad (2.10)$$

is invariant under a SL(4,R) transformation and equal for both the $G$ group and the $H$ subgroup. There is a representation of $A$ on the bundle $S$ corresponding to the induced representation. The invariance (equality) of the product $JJ$ must be valid in any representation of $A$, although the value of the product may differ from one representation to the other. For the induced function representations valued in the sl(2,C) algebra (Pauli matrices), the product becomes

$$\int J_\mu^{\hat{\alpha}}(k_1,k_2) \kappa_0 \kappa_{\hat{\alpha}} J_{\hat{\beta}}^\mu(k_2,k_3) \kappa^0 \kappa^{\hat{\beta}} dk_2 \equiv F(k_1,k_3) = F_0 = $$

$$\int J_\mu^{\hat{\alpha}}(k'_1,k'_2) \kappa_0 \kappa_{\hat{\alpha}} J_{\hat{\beta}}^{\mu'}(k'_2,k'_3) \kappa^0 \kappa^{\hat{\beta}} dk'_2 \quad , \qquad (2.11)$$

which, must be a constant 4×4 matrix, independent of $k_1$ and $k_3$.

The expression for the mass becomes,

$$m = \frac{m_g}{4V(A_R)} \mathrm{tr} \int_{K_R} F(k,k) dk \quad . \qquad (2.12)$$

The integrand is the same $F_0$ constant for both groups, but the range of integration $X$ differs. Integration is on a subspace $K_R \subset K$ of relativistic inequivalent points of $K$ for the group $G$ and on a subspace $C_R \subset C \subset K$ for the group $H$. The expressions for the masses corresponding to $G$-excitations and $H$-excitations become,

$$m_G = \frac{m_g}{4V(A_R)} \mathrm{tr} \int_{K_R} F(k,k) dk = \frac{V(K_R)}{4V(A_R)} m_g \, \mathrm{tr}(F_0) \; , \qquad (2.13)$$

$$m_H = \frac{m_g}{4V(A_R)} \mathrm{tr} \int_{C_R} F(k,k) dk = \frac{m_g}{4V(A_R)} \mathrm{tr}(F_0) V(C_R) = \frac{m_G V(C_R)}{V(K_R)} \quad . \qquad (2.14)$$

The bare mass parameters $m$ are related to integration on certain subspaces of the coset spaces, depending proportionally on the volumes of the respective coset spaces. In other words, the ratio of the bare mass parameters for representations of SL(4,R) and Sp(4,R), induced from the same SL(2,C) representation as functions on cosets $K$ and $C \subset K$ would be equal to the ratio of the volumes of the respective subspaces.

## 3. Volume Ratios.

### 3.1. Volume of C Space

First consider, the volume of the four dimensional symmetric space Sp(4,R)/SL(2,C) which coincides with the quotient SO(3,2)/SO(3,1) as shown in appendix B. In the regular representation, it has the structure

$$C = \begin{bmatrix} \begin{bmatrix} & * & \end{bmatrix} & \begin{bmatrix} x^0 \\ x^1 \\ x^2 \\ x^3 \end{bmatrix} \\ \begin{bmatrix} & * & \end{bmatrix} & [x^4] \end{bmatrix} , \qquad (3.1)$$

where $x^4$ must be a function of the $x^\mu$ imposed by the group structure [16],

$$x^4 = \left(1 - \eta_{\mu\nu} x^\mu x^\nu\right)^{\frac{1}{2}} . \qquad (3.2)$$

The Euclidian volume element $dV(c)$ given in terms of the forms $dx^\mu(c)$ varies over the four dimensional coset. The invariant measure $d\mu(c)$ is determined by weighing the Euclidian element by a density equal to the inverse of the Jacobian of the transformation generated by a translation in the coset,

$$d\mu(c) = \frac{dV(c)}{\|J(c)\|} , \qquad (3.3)$$

$$d\mu = \frac{dx^0 \wedge dx^1 \wedge dx^2 \wedge dx^3}{\sqrt{1 - \eta_{\mu\nu} x^\mu x^\nu}} = \frac{dV}{\sqrt{1 - \eta_{\mu\nu} x^\mu x^\nu}} , \qquad (3.4)$$

or in $R^5$ with $w$ as the fifth coordinate $x^4$,

$$\delta\left(\sqrt{w^2 + \eta_{\mu\nu} x^\mu x^\nu} - 1\right) dx^0 \wedge dx^1 \wedge dx^2 \wedge dx^3 \wedge dw , \qquad (3.5)$$

leading to

$$d\mu = \frac{dx^0 dx^1 dx^2 dx^3}{|w|} = \frac{dV}{|w|} , \qquad (3.6)$$

where,

$$\eta_{\mu\nu} x^\mu x^\nu \rightarrow \eta_{\mu\nu} x^\mu x^\nu + w^2 . \qquad (3.7)$$

Inspection of this equation allows us to physically interpret the parameter $w$ as a measure of a variation of mass energy throughout the coset $C$,

$$|w| = \sqrt{1 - \eta_{\mu\nu} x^\mu x^\nu} = \sqrt{1 - m^2} . \qquad (3.8)$$

The presence of the Jacobian means that that we should use coordinates adapted to this symmetric space, polar hyperbolic coordinates, to calculate its invariant volume density,

$$(C) = \int_C \frac{d\mu}{\pm |w|} = \int_C \sqrt{-g} \, dV . \qquad (3.9)$$

The quotient space must be one of the standard four dimensional hyperboloids $H^4$ [17]. We add a superscript $\alpha$ that indicates the number of negative signs in the invariant standard definition of $H^{n,\alpha}$,

$$1 = w^2 + \sum_{i=\alpha+1}^{n}(x_i)^2 - \sum_{i=1}^{\alpha}(x_i)^2 \quad . \tag{3.10}$$

In particular, the space corresponds to the hyperboloid $H^{4,3}$,

$$1 = w^2 + u^2 - x^2 - y^2 - z^2 = w^2 + u^2 - k^2 \quad , \tag{3.11}$$

in terms of the coordinates $u, k$ corresponding respectivelly to the energy and to the momentum absolute value, and an overall parameter $\lambda$ that characterizes the size of a particular four dimensional hyperboloid. We introduce a parametrization in terms of arcs in the symmetric space by defining the hyperbolic coordinates $\varphi, \theta, \beta, \chi$,

$$\cos\varphi = \frac{x}{\sqrt{x^2 + y^2}} \quad , \tag{3.12}$$

$$\cos\theta = \frac{z}{k} \quad , \tag{3.13}$$

$$\cosh\beta = \frac{u}{\sqrt{u^2 - k^2}} \quad , \tag{3.14}$$

$$\cos\chi = \frac{w}{\lambda} \quad . \tag{3.15}$$

It should be noted that the hyperbolic arc parameter $\beta$ is not the relativistic velocity but is related to it by

$$\tanh\beta = \frac{k}{u} = \frac{v}{c} \quad . \tag{3.16}$$

In particular the volume of $C$ is obtained by an integration over this Minkowskian momentum space, where the coordinates $x$ stand for $u, k$. We split the integration into the angular integration on the compact sphere $S^2$, the radial boost boost $\beta$ and the energy parameter $\chi$.,

$$\begin{aligned}V(C) &= \int_0^\pi d\chi \sin^3\chi\psi \int_0^\beta d\beta \sinh^2\beta \int_0^{4\pi} d^2\Omega = \\ &\quad \frac{16\pi}{3}\int_0^\beta d\beta\sinh^2\beta = \frac{16\pi}{3}I_C(\beta) \quad ,\end{aligned} \tag{3.17}$$

in terms of a boost integral $I(\beta)$.

### 3.2. Volume of K space.

For the volume of $K$, the integration is over an 8 dimensional symmetric space. This space $G/G_+$ has a complex structure and is a non Hermitian space. The center of $G_+$, which is not discrete, contains a generating element $\kappa_5$ whose square is -1. We shall designate by $2J$ the restriction of $ad(\kappa_5)$ to the space $TK_k$. This space, that has for base the 8 matrices $\kappa_\alpha \kappa_\beta \kappa_5$, is the proper subspace corresponding to the eigen value -1 of the operator $J^2$,

$$J^2(x^\lambda\kappa_\lambda + y^\lambda\kappa_\lambda\kappa_5) = \tfrac{1}{4}[\kappa_5,[\kappa_5,x^\lambda\kappa_\lambda + y^\lambda\kappa_\lambda\kappa_5]] = \\ -x^\lambda\kappa_\lambda - y^\lambda\kappa_\lambda\kappa_5 \quad . \tag{3.18}$$

The endomorphism $J$ defines an almost complex structure over $K$. In addition, using the Killing metric, in the Clifford representation,

$$g(Ja, Jb) = \tfrac{1}{4}\mathrm{tr}(JaJb) = \tfrac{1}{4}\mathrm{tr}(J^2(-a)b) = \tfrac{1}{4}\mathrm{tr}(ab) = g(a,b) \quad , \tag{3.19}$$

we have that the complex structure preserves the pseudo Riemannian (Minkowskian) Killing metric. Furthermore the torsion $S$ is zero,

$$S(a,b) = [a,b] + J[Ja,b] + J[a,Jb] - [Ja,Jb] = 0 \ . \quad (3.20)$$

In this form, the conditions for $J$ to be an integrable complex structure, invariant by $G$, are met and the space $K$ is a non Hermitian complex symmetric space [18].

It is known that the Hermitian symmetric spaces are classified by certain group quotients. The symmetric space $K$ is a non compact real form of the complex symmetric space corresponding to the complex extension of the non compact group SU(2,2) and its quotients as shown in appendix B. This space coincides with the quotient SO(3,3)/SO(3,1)⊗SO(2) of the SO(4,2) series. In particular we have the 8 dimensional spaces

$$R \equiv \frac{SO(4,2)}{SO(4) \times SO(2)} \approx \frac{SL(4,R)}{SL(2,C) \times SO(2)} \cong K \cong \cdots \approx \frac{SO(6)}{SO(4) \times SO(2)} \ . \quad (3.21)$$

which are the five real forms characterized by SO(4,2). The extreme spaces correspond to the two Hermitian spaces, compact and non compact $R$.

Between the extremes we find the three non Hermitian non compact spaces, in particular the space of interest $K$. In the regular representation these quotients have the matricial structure,

$$K = \begin{bmatrix} \begin{bmatrix} & & \\ & * & \\ & & \end{bmatrix} & \begin{bmatrix} x^0 & y^0 \\ x^1 & y^1 \\ x^2 & y^2 \\ x^3 & y^3 \end{bmatrix} \\ \begin{bmatrix} & * & \end{bmatrix} & \begin{bmatrix} x^4 & y^4 \\ x^5 & y^5 \end{bmatrix} \end{bmatrix}, \quad (3.22)$$

where the lower right submatrix determines the conditions,

$$\begin{bmatrix} x^4 & y^4 \\ x^5 & y^5 \end{bmatrix} = \begin{bmatrix} 1 + x \bullet x & x \bullet y \\ y \bullet x & 1 + y \bullet y \end{bmatrix}^{\frac{1}{2}}, \quad (3.23)$$

imposed by the corresponding associated groups on the coordinates $x^4$, $x^5$, $y^4$, $y^5$ in higher dimensional spaces ($d>8$), expressed by the scalar product in this submatrix, in terms of the corresponding 4-vectors $x, y$ and respective metric, related to the Euclidian metric by Weyl's unitary trick. As in the previous case, this condition determines a unit symmetric space.

Since these conditions are difficult to analyze, it is convenient to find the volume using the complex structure of the manifold. We may introduce complex coordinates $z^\mu$ on $K$. In this way we obtain a symmetric bilinear complex metric in the complex four dimensional space $K$,

$$\frac{1}{4} \text{tr}(z^\alpha \kappa_\alpha)^2 = -z^\alpha z^\beta \eta_{\alpha\beta} \in C \ , \quad (3.24)$$

$$z^\mu = |z^\mu| e^{i\psi^\mu} \quad (3.25)$$

The group that preserves this symmetric bilinear complex metric is the orthogonal group SO(4,C),

There are two standard projections of the complex numbers $C$ on the real numbers $R$, the real part and the modulus. In a similar manner as the modulus projects the complex plane to the real half line, we define an equivalence relation $S$ in the points on $K$ by defining equivalent points as points with equal moduli coordinates. This equivalence relation may be expressed by

$$S = S^1 \times S^1 \times S^1 \times S^1 , \quad (3.26)$$

where $S^1$ is the one dimensional phase sphere. We may define the 4 dimensional quotient $Q$ by

$$Q = \frac{K}{S} \ . \quad (3.27)$$

It is then convenient to parametrize $K$ in accordance with this equivalence relation $S$. Each of the 4 complex coordinates $z$ has a modulus $|z|$ and a phase $\psi$ which will be used as parameters. The Haar measure in terms of the Euclidian measure should correspond to integration on a complex 4 dimensional space. We choose the parameters defining the Euclidian volume element at the identity,

$$dV(I) = d|z^0|(I) \wedge d\psi^0(I) \wedge d|z^1|(I) \wedge d\psi(I)^1 \wedge \\ d|z^2|(I) \wedge d\psi^2(I) \wedge d|z^3|(I) \wedge d\psi^3(I) \quad , \tag{3.28}$$

and translate it to the point $k$ by a coset operation,

$$d\mu(k) = \frac{dV(k)}{\|J(k)\|} \quad . \tag{3.29}$$

As before, the Jacobian is a measure of a variation of mass energy throughout the coset $K$. We calculate the volume integral using the invariant volume density,

$$d\mu = \frac{d|z^0| \wedge d|z^1| \wedge d|z^2| \wedge d|z^3| \wedge d\psi^0 \wedge d\psi^1 \wedge d\psi^2 \wedge d\psi^3}{\|J(k)\|} \quad . \tag{3.30}$$

The Jacobian does not depend on the phases and the volume element of $S$ is separable from $d\mu$ defining a complementary volume element corresponding to the quotient $Q$. The volume of $K$ will then be the product of the volumes of $S$ and $Q$, as indicated also by eq. (3.2.27). It may be seen that $Q$ is a symmetric space, by taking as its representative points those with real coordinates. On the other hand $Q$ must be non compact, otherwise $K$ would also be compact since the relation $S$ is compact. The quotient cannot be a product of a manifold times a discrete set because then $K$ would have disconnected components. The SO(4,R) subgroup of SO(4,C), which acts on this subspace $Q$ of $K$ preserving the bilinear metric in this space, has $S^3$ as orbit. Therefore $Q$ is symmetric, four dimensional, non compact with a three dimensional compact subspace $\Sigma_S$ equal to the Riemannian sphere $S^3$,

$$Q = \frac{K}{S} \supset S^3 \quad . \tag{3.31}$$

The quotient $Q$ must, then, be the hyperboloid $H^{4,4}$, the only one with a $S^3$ spacelike subspace, characterized in five dimensional space by the invariant,

$$1 = -w^2 + u^2 - x^2 - y^2 - z^2 = -w^2 + u^2 - k^2 \quad . \tag{3.32}$$

The physical boost parameter $\beta$ (the one that represents a change in kinetic energy) is the hyperbolic arc parameter along the orbit of a one parameter non compact subgroup. The parametrization is in terms of the hyperbolic coordinates $\varphi, \theta, \zeta, \beta$ defined, instead of eqs. (3.14, 3.15), by,

$$\cosh\beta = \frac{u}{\lambda} \quad , \tag{3.33}$$

$$\cos\zeta = \frac{w}{\sqrt{w^2 + k^2}} \quad . \tag{3.34}$$

We split the integration into the integration on the 4 compact phase spaces $S$ and integration on the 4 dimensional space $Q$, parametrized by the coordinate moduli. The last integration is further split into integration on the compact 3 spheres $S^3$ and integration on the complementary non compact direction, which corresponds to boosts $\beta$, obtaining,

$$V(K) = \int_K \sqrt{-g}\, d^4V d\psi_0 d\psi_1 d\psi_2 d\psi_3 = V(Q) \times V(S) = \\ \int_0^\beta d\beta \sinh^3\beta \int_0^\pi d\zeta \sin^2\zeta \int_0^{4\pi} d^3\Omega \times \left(\int_0^{2\pi} d\psi\right)^4 \quad , \tag{3.35}$$

$$V(K) = (2\pi^2)(2\pi)^4 \int_0^\beta \sinh^3 \beta d\beta = 2^5 \pi^6 I_K(\beta) \quad , \tag{3.36}$$

in terms of another boost integral $I(\beta)$.

### 3.3 Ratio of Geometric Volumes

We expect that the ratio of the volumes $V$ of the inequivalent subspaces corresponding to fundamental representations of spin *1/2* should be related to the ratio of the corresponding physical bare masses. As indicated in section 2, there is a subset of points of the symmetric spaces that are relativistic equivalent. We have to eliminate these equivalent points by dividing by the equivalence relation $R$ under the boosts of SO(3,1). Equivalent points are related by a Lorentz boost transformation of magnitude $\beta$. There are as many equivalent points as the volume of the orbit developed by the parameter $\beta$.

The respective inequivalent volumes are, for *C*,

$$V(C_R) = \frac{V(C)}{V(R(\beta))} = \frac{\frac{16\pi}{3} I_C(\beta)}{I_C(\beta)} = \frac{16\pi}{3} \quad , \tag{3.37}$$

and for *K*,

$$V(K_R) = \frac{V(K)}{V(R(\beta))} = \frac{2^5 \pi^6 I_K(\beta)}{I_K(\beta)} = 2^5 \pi^6 \quad . \tag{3.38}$$

Although the volumes of the non compact spaces diverge, their ratio taking in consideration the equivalence relation $R$ has a well defined limit, obtaining,

$$\frac{V(K_R)}{V(C_R)} = 6\pi^5 \quad . \tag{3.39}$$

We actually have shown in this section a theorem that says: *The ratio of the volumes of K and C, up to the equivalence relation R under the relativity isotropy subgroup, is finite and has the value $6\pi^5$*.

## 4. Physical Mass Ratios.

It is clear from the discussion in section 1 that, for a constant solution, the masses are proportional to the respective volumes $V(K_R)$. The constants of proportionality only depend on the specified inducing SL(2,C) representations. In particular for any two representation induced from the the spin 1/2 representation of *L*, the respective constants are equal.

Armed with the theorem of the previous section, the ratio of masses of *G*-excitations and *H*-excitations given by the fundamental representations of the groups *G* and its subgroup *H* induced from the spin 1/2 representation of *L*, for the constant solution, equals the ratio of the volumes of the inequivalent subspaces of the respective cosets $G/L_1$ and $H/L$. The ratio of bare masses for these geometric excitations has the finite exact value,

$$\frac{m_G}{m_H} = \frac{V(K_R)}{V(C_R)} = 6\pi^5 = 1836.1181 \approx \frac{m_p}{m_e} \quad , \tag{4.1}$$

which is a very good approximation for the ratio of the experimental physical values for the proton mass and the electron mass. This geometrical expression was previously known but not physically explained [19,20,21].

If we reject the easy explanation that this result is a pure coincidence, we should try to relate the *G* group to the proton and the *H* group to the electron. If this were the case the only other dynamical subgroup *L*=SL(2,C) of *G* should lead to a similar mass ratio. Previously we have related *L* to the neutrino. In this case the quotient space is the identity and we get,

$$\frac{m_L}{m_H} = \frac{V_R(L/L)}{V_R(C)} = \frac{V_R(I)}{V_R(C)} = 0 = \frac{m_\nu}{m_e} \quad , \tag{4.2}$$

which is in accordance with the zero bare mass of the neutrino. There may be small corrections to obtain the physical masses due to excitation contributions.

## 5. Conclusion.

The ratio of geometrical masses $m_L/m_H$, given by the volumes of the quotients SL(4,R)/SL$_1$(2,C) and Sp(4,R)/SL(2,C)) up to the equivalence relation under SL(2,C) is equal, with a discrepancy of $2\times10^{-5}$, to the value of the ratio of the masses of the proton and the electron.

It appears that the values of the bare masses of the proton and electron may be calculated as masses of geometric excitations in this unified theory. These values necessarily need correction terms to obtain the physical masses, because of the interactions of the excitations. These corrections, due to the self interaction of the excitations may be expected to be of the order of $\alpha^2$, equal to the order of the discrepancy.

## 6. Appendix A.

The non linear equations of the theory are applicable to an isolated background physical system interacting with itself. Of course the equations must be expressed in terms of components with respect to an arbitrary reference frame. A reference frame adapted to an arbitrary observer introduces arbitrary fields which do not contain any information related to the physical system in question.

Any excitation must be associated to a definite background. An arbitrary observation of an excitation property depends on both the excitation and the background, but the physical observer must be the same for both excitation and background. We may use the freedom to select the reference frame, to refer the excitation to the physical frame defined by its own background.

We have chosen the current density 3-form $J$ corresponding to the vector,

$$J^\mu = \tilde{e}\,\kappa^{\hat{\alpha}} u_{\hat{\alpha}}^\mu e \quad , \tag{6.1}$$

in terms of the matter spinor frame $e$ and the orthonormal space-time tetrad $u$.

Since we selected that the background be referred to itself, the background matter frame $e_b$, referred to $e_r$ becomes the group identity $I$. Actually this generalizes comoving coordinates (coordinates adapted to dust matter geodesics) [22]. We adopt coordinates adapted to local background matter frames (the only non arbitrary frame is itself, as are the comoving coordinates defined by an isolated fluid). If the frame $e$ becomes the identity $I$ (in the group), the background current density becomes a constant, invariant under the action of the connections,

$$J = \kappa_\alpha dx^\alpha \quad . \tag{6.2}$$

Comparison of an object with itself gives trivial information. For example free matter or an observer are always at rest with themselves, no velocity, no acceleration, no self forces, etc. In its own reference frame these effects actually disappear. Only constant self energy terms, determined by the non linearity, makes sense and should be the origin of the constant mass.

In this reference frame the background field equation,

$$D^* \Omega_b = 4\pi\alpha\, {}^*\! J_b \quad , \tag{6.3}$$

admits a constant $\omega$ solution. The differential expression for the left side of this equation reduces to an algebraic expression, a triple product of $\omega$, giving,

$$\omega \wedge {}^*(\omega \wedge \omega) - {}^*(\omega \wedge \omega) \wedge \omega = 4\pi\alpha \kappa_\alpha dx^\alpha \quad , \tag{6.4}$$

which has for a trivial particular solution,

$$\omega = -m_g dx^{\hat{\alpha}} \kappa_{\hat{\alpha}} \quad , \tag{6.5}$$

where $m_g$ is a geometric dimensionless constant

$$m_g = \left(\pi\alpha/3\right)^{1/3} \quad . \tag{6.6}$$

In an arbitrary reference system the connection becomes, in terms of the current form $J$,

$$\omega = e^{-1}\left(-m_g dx^\alpha \kappa_\alpha\right)e + e^{-1}de = -m_g J + e^{-1}de. \qquad (6.7)$$

The trivial connection is essentially proportional to the current, up to an automorphism. It should be noted that in the expression for $\omega$, the term containing the current $J$ is the potential tensorial form $\Lambda$, used in chapter 8 in the definition of mass. Its subtraction from $\omega$ gives an object, $e^{-1}de$, that transforms as a connection.

## 7. Appendix B.

The series of quotient spaces related to the $A_3$ Cartan space are of interest. In particular we choose the involutive automorphism of type AIII(p=2,q=2) that determines a seven dimensional compact subgroup $G_+$. We obtain, in this manner, a series of eight dimensional spaces, characterized by the non compact group SU(2,2), corresponding to the Riemannian space $G/G_+$ and its dual $G^*/G_+$,

$$\frac{SU(4)}{SU(2)\otimes SU(2)\otimes U(1)} \approx \frac{SU^*(4)}{SL(2,C)\otimes SO(2)} \approx \frac{SU(2,2)}{SL(2,C)\otimes SO(1,1)} \approx$$
$$\approx \frac{SL(4,R)}{SL(2,C)\otimes SO(2)} \approx \frac{SU(2,2)}{SU(2)\otimes SU(2)\otimes U(1)} \qquad . \quad (7.1)$$

Due to the isomorphism of the spaces $A_3$ and $D_3$ we have the isomorphic series, characterized by the non compact group SO(4,2), corresponding to the Riemannian space $G/G_+$ and its dual $G^*/G_+$, with involution of the type BDI(p=4,q=2),

$$\frac{SO(6)}{SO(4)\otimes SO(2)} \approx \frac{SO(5,1)}{SO(3,1)\otimes SO(2)} \approx \frac{SO(4,2)}{SO(3,1)\otimes SO(1,1)}$$
$$\approx \frac{SO(3,3)}{SO(3,1)\otimes SO(2)} \approx \frac{SO(4,2)}{SO(4)\otimes SO(2)} \qquad . \quad (7.2)$$

The characters of the real forms of both isomorphic series are -8, -4, 0, +4, +8.

Another series of interest are the ones related to the $C_2$ Cartan space. In particular, we choose the involutive automorphism of the type CII(p=2,q=2) that determines a six dimensional compact subgroup. We obtain a series of four dimensional spaces, characterized by the non compact group USp(2,2), corresponding to the Riemannian space $G/G_+$ and its dual $G^*/G_+$,

$$\frac{USp(4)}{USp(2)\otimes USp(2)} \approx \frac{USp(2,2)}{Sp(2,C)} \approx \frac{Sp(4,R)}{Sp(2,R)\otimes Sp(2,R)} \approx$$
$$\approx \frac{Sp(4,R)}{Sp(2,C)} \approx \frac{USp(2,2)}{USp(2)\otimes USp(2)} \qquad . \quad (7.3)$$

Also, due to the isomorphism of the spaces $B_2$ and $C_2$ we have the isomorphic series, characterized by the non compact group SO(4,1), corresponding to the Riemannian space $G/G_+$ and its dual $G^*/G_+$ with involution of the type BDI(p=4,q=1),

$$\frac{SO(5)}{SO(4)} \approx \frac{SO(4,1)}{SO(3,1)} \approx \frac{SO(3,2)}{SO(2,2)} \approx \frac{SO(3,2)}{SO(3,1)} \approx \frac{SO(4,1)}{SO(4)} \qquad . \quad (7.4)$$

The characters of the real forms of both isomorphic series are -4, -2, 0, +2, +4.